\documentclass[]{emulateapj}

\usepackage{epsfig,amsmath,amssymb,amsfonts,mathrsfs,latexsym,graphicx,lscape}
\usepackage{hyperref}

\begin{document}

\title{The faint ``heartbeats'' of IGR~J17091--3624: an exceptional
  black-hole candidate}

\author{D. Altamirano\altaffilmark{1},
T. Belloni\altaffilmark{2},
M. Linares\altaffilmark{3}, 
M. van der Klis\altaffilmark{1},
R. Wijnands\altaffilmark{1}, 
P. A. Curran\altaffilmark{4},
M. Kalamkar\altaffilmark{1},
H. Stiele\altaffilmark{2},
S. Motta\altaffilmark{2},
T. Mu\~noz-Darias\altaffilmark{2,5}, 
P. Casella\altaffilmark{6},
H. Krimm\altaffilmark{7}}

\altaffiltext{1}{Email: d.altamirano@uva.nl ; Astronomical Institute,
  ``Anton Pannekoek'', University of Amsterdam, Science Park 904,
  1098XH, Amsterdam, The Netherlands}

\altaffiltext{2}{INAF-Osservatorio Astronomico di Brera, Via
  E. Bianchi 46, I-23807 Merate (LC), Italy}

\altaffiltext{3}{Massachusetts Institute of Technology - Kavli
Institute for Astrophysics and Space Research, Cambridge, MA 02139,
USA}

\altaffiltext{4}{Laboratoire AIM, CEA DSM/IRFU/SAp, Centre de Saclay,
  F-91191 Gif-sur-Yvette, France}

\altaffiltext{5}{Instituto de Astrof\'{\i}sica de Canarias, 38200 La Laguna, Tenerife, Spain}

\altaffiltext{6}{School of Physics and Astronomy, University of
  Southampton, Southampton, Hampshire, SO17 1BJ, United Kingdom}

\altaffiltext{7}{CRESST and NASA Goddard Space Flight Center, Greenbelt, MD 20771, USA ; Universities Space Research Association, 10211 Wincopin Circle, Suite 500, Columbia, MD 21044, USA}

\begin{abstract}

  We report on the first 180 days of RXTE observations of the outburst
  of the black hole candidate IGR~J17091--3624.
  This source exhibits a broad variety of complex light curve patterns
  including periods of strong flares alternating with quiet intervals.
  Similar patterns in the X-ray light curves have been seen in the (up
  to now) unique black hole system GRS~1915+105.
  In the context of the variability classes defined by
  \citet{Belloni00} for GRS~1915+105, we find that IGR~J17091--3624
  shows the $\nu$, $\rho$, $\alpha$, $\lambda$, $\beta$ and $\mu$
  classes as well as quiet periods which resemble the $\chi$ class,
  all occurring at 2-60 keV count rate levels which can be 10-50
  times lower than observed in GRS~1915+105.
  The so-called $\rho$ class ``heartbeats'' occur as fast as every few
  seconds and as slow as $\sim$100 seconds, tracing a loop in the
  hardness-intensity diagram which resembles that previously seen in
  GRS~1915+105.
  However, while GRS~1915+105 traverses this loop clockwise,
  IGR~J17091--3624 does so in the opposite sense.
  We briefly discuss our findings in the context of the models
  proposed for GRS~1915+105 and find that either all models requiring
  near Eddington luminosities for GRS~1915+105-like variability fail,
  or IGR~J17091--3624 lies at a distance well in excess of 20 kpc or,
  it harbors one of the least massive black holes known ($< 3
  M_\odot$).

\end{abstract}

\section{Introduction}\label{sec:intro}

Observations with the \textit{Rossi X-ray Timing Explorer} (RXTE) have
led to extraordinary progress in the knowledge of the variability
properties of many different types of sources, particularly of black
hole candidates (BHCs) and neutron stars (NSs) in low-mass X-ray
binaries \citep[e.g,][]{Vanderklis06}.
Both BHCs and NS are known to exhibit distinct ``accretion states''
(usually defined in terms of their X-ray spectral shape and
variability), whose characteristics are thought to be intimately
related to the physics of the accretion flow and its interaction with
the central compact object \citep[e.g.,][]{Belloni10b, Remillard06,
  Vanderklis06,Belloni11}.

Of all Galactic BHC X-ray binaries known today, one of the most
prolific in terms of state transitions is GRS~1915+105.
In outburst since its discovery in 1992 with WATCH
\citep{Castro-Tirado92}, it is a 33 days orbital period binary system
\citep{Greiner01} harboring a $14\pm4.4 \ M_{\odot}$ black hole
\citep{Greiner01,Harlaftis04}. At a distance of $\sim12.5$ kpc
\citep{Mirabel94}, GRS~1915+105 is very often at Eddington or
super-Eddington luminosity \citep[e.g., ][]{Done04}.

\begin{figure*} 
\centering
\resizebox{2\columnwidth}{!}{\rotatebox{-90}{\includegraphics[clip]{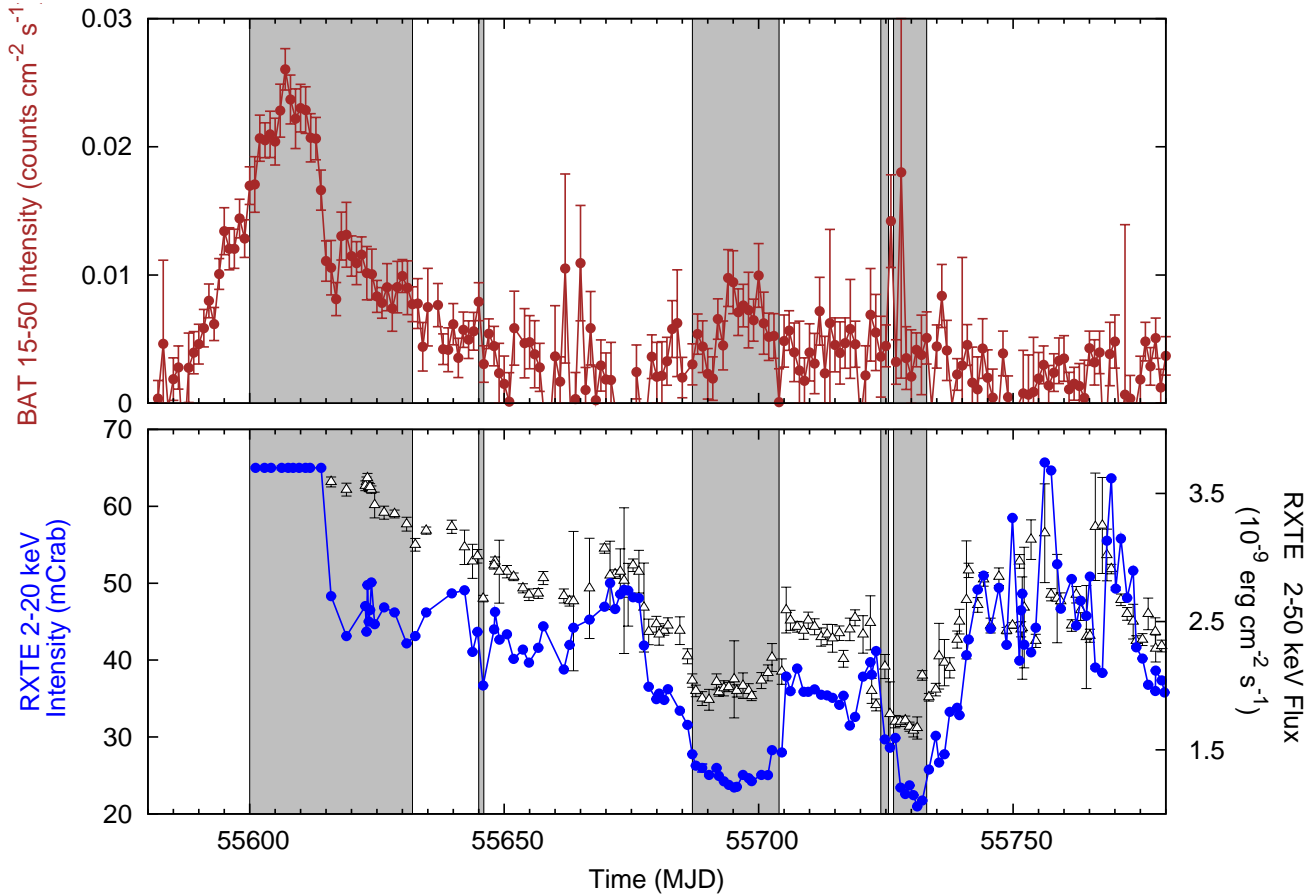}}}
\caption{\textit{Upper Panel}: \textit{Swift/BAT} light curve. \textit{Lower
  Panel}: Circles: RXTE Crab normalized 2-20 keV intensity.  RXTE
values are averaged per observation, background subtracted and
dead-time corrected.  Triangles: RXTE 2-50 keV flux RXTE observations
between day 55601 and 55615 are affected by the nearby source GX~349+2
(Sco X-2); the intensity during this period was arbitrarily fixed to
65 mCrab and no X-ray flux is reported. Shaded areas mark quiet
periods (see text). }
\label{fig:outburstlc}
\end{figure*}

GRS~1915+105 is known to show quasi-periodic oscillations (QPOs) in
the 0.001--70 Hz range, some of which are the same as those generally
seen in other BHCs \citep[]{Morgan97, Markwardt99, Reig00a,
  Strohmayer01b, Soleri08}. However, so far GRS~1915+105 has been
unique in that its X-ray light curves exhibit more than a dozen
different patterns of variability usually called ``classes'' (which are
referred to with Greek letters), most of which are high amplitude and
highly-structured \citep[e.g.,][]{Belloni00}.
Most of this structured variability is thought to be due to limit
cycles of accretion and ejection in an unstable disk \citep[see, e.g.,
][and references within]{Belloni97, Mirabel98, Tagger04, Neilsen11}.
Much effort has been expended in order to understand why GRS~1915+105
is so unusual among BHCs \citep[e.g.,][]{Fender04}. It has been
proposed that the high accretion rate estimated for GRS~1915+105 might
be the determining factor \citep[e.g.,][]{Done04}; however, the lack
of any other source showing similar characteristics has prevented
definite conclusions.

\begin{figure} 
\centering
\resizebox{1\columnwidth}{!}{\rotatebox{0}{\includegraphics[clip]{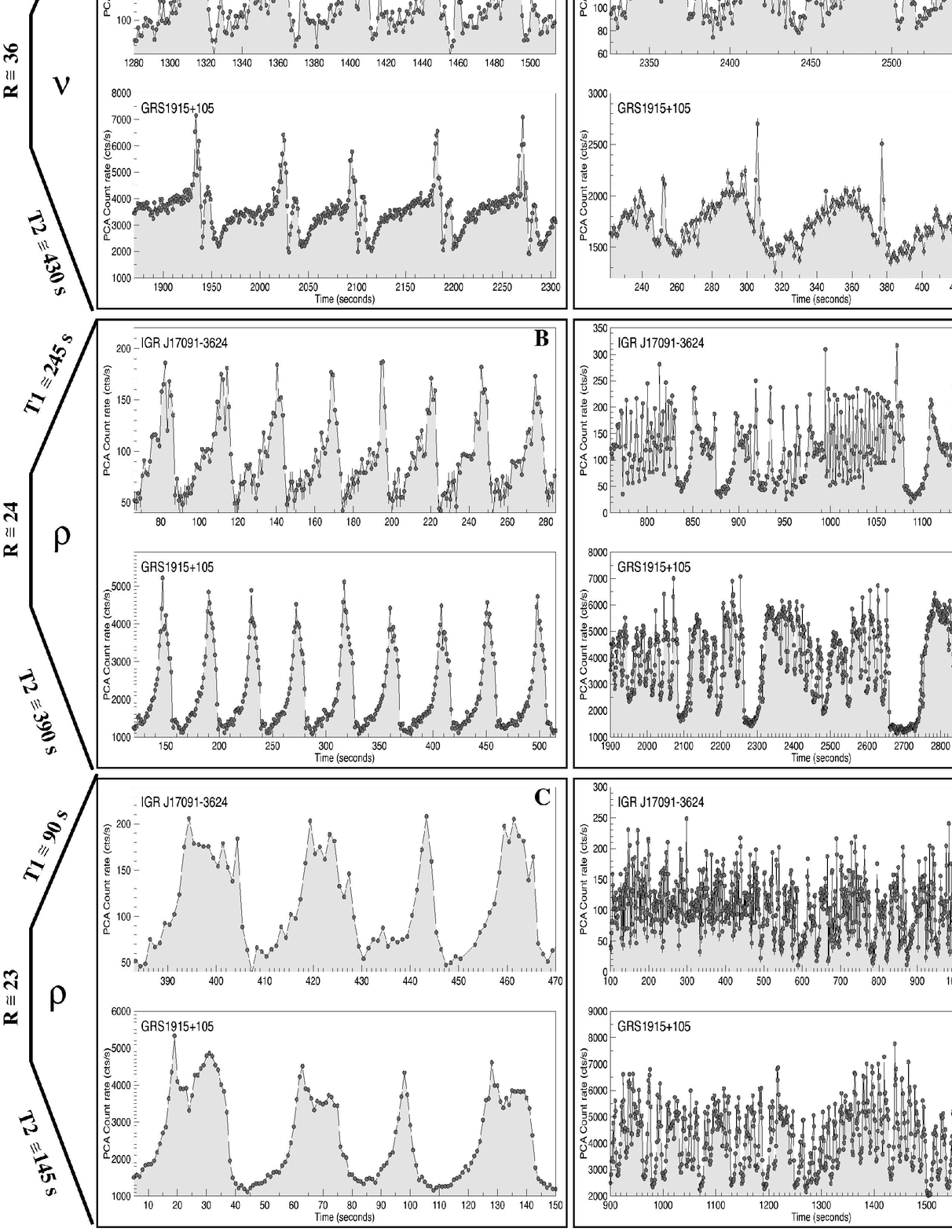}}}
\caption{The upper and lower panels of each frame show a light curve
  of IGR~J17091--3624 and GRS~1915+105, respectively. Count rates are
  given in 1 s bins, per PCU, 2-60 keV and background subtracted. The
  IGR~J17091--3624 A-C intervals last $T1$ seconds and come from
  observations 96420-01-05-00, -06-00 and -07-01, respectively.  For
  GRS~1915+105 they last $T2$ seconds and come from observations
  10408-01-40-00, 20402-01-34-00 and 93701-01-02-01, respectively. $R
  \equiv MIN_{GRS~1915+105} / MIN_{IGR~J17091-3624}$, where $MIN$ is
  an average count rate during minima in the above light curve and is
  given for each frame as an order of magnitude estimate of the flux
  ratio. Greek letters on the side indicate variability the class. }
\label{fig:complc1}
\end{figure}

IGR~J17091--3624 was discovered with INTEGRAL/IBIS during a Galactic
Center observation on April 2003 \citep{Kuulkers03b} and later again
in 2007 \citep{Capitanio09}. Reexamination of archival data from
different missions showed that IGR~J17091--3624 was also active in
1994, 1996, 2001 \citep{Revnivtsev03a, Intzand03, Capitanio06}.
Although a neutron star could not be excluded based on spectral
characteristics and the outburst radio/X-ray flux ratio,
\citet{Capitanio06} concluded that IGR~J17091--3624 is most probably a
black hole. 
A new outburst was detected with \textit{Swift/BAT} in February 2011
\citep{Krimm11}; the radio/X-ray characteristics during the first
$\sim$40 days, combined with the discovery of QPOs
\citep{Rodriguez11a}, further suggested that IGR~J17091--3624 is a
black hole.

Recently we reported the discovery of 10 mHz QPOs in RXTE observations
of IGR~J17091--3624 \citep{Altamirano11a} very similar to those in
GRS~1915+105. The suggested link between these sources was
strengthened by our discovery of
(i) a continuous progression of regular, quasi-periodic flares
occurring at a rate of 25--30 mHz \citep{Altamirano11b}, and
(ii) a broad variety of complex patterns alternating with quiet
intervals \citep{Altamirano11b,Altamirano11c}, 
resembling respectively, the so called $\rho$ class (``heartbeat'')
oscillations and the complex $\beta$ class patterns, both so far seen
only in GRS~1915+105.

The existence of a source showing similar X-ray variability to that
seen in GRS~1915+105 opens a new window of opportunities to understand
the physical mechanism that produces the highly structured X-ray
variability.
Multi-wavelength and high spectral resolution studies of
IGR~J17091--3624 as compared with GRS~1915+105 can help gaining
further insights into the role of disc-jet coupling \citep[e.g.,
review by][]{Fender04} and accretion disk winds \citep[e.g,
][]{Neilsen11} in accreting BHCs.
Furthermore, it allows the possibility to test the role of the
accretion disk size, evolutionary state of the companion star, and the
evolution of the disc structure as an explanation of the long- and
short-term X-ray variability seen in
GRS~1915+105 \citep[e.g.,][]{Done04}.

In this Letter we use RXTE data to describe the phenomenological
similarities between IGR~J17091--3624 and GRS~1915+105.
We demonstrate that IGR~J17091--3624 shows the same type of high
amplitude and highly-structured variability previously observed only
in GRS~1915+105, although at much lower count rates.
The analysis presented here is the first step of a larger program to
compare IGR~J17091--3624 and GRS~1915+105 in detail using data
obtained with \textit{RXTE}, \textit{Swift}, \textit{XMM-Newton} as
well as optical facilities.

\begin{figure} 
\centering
\resizebox{1\columnwidth}{!}{\rotatebox{0}{\includegraphics[clip]{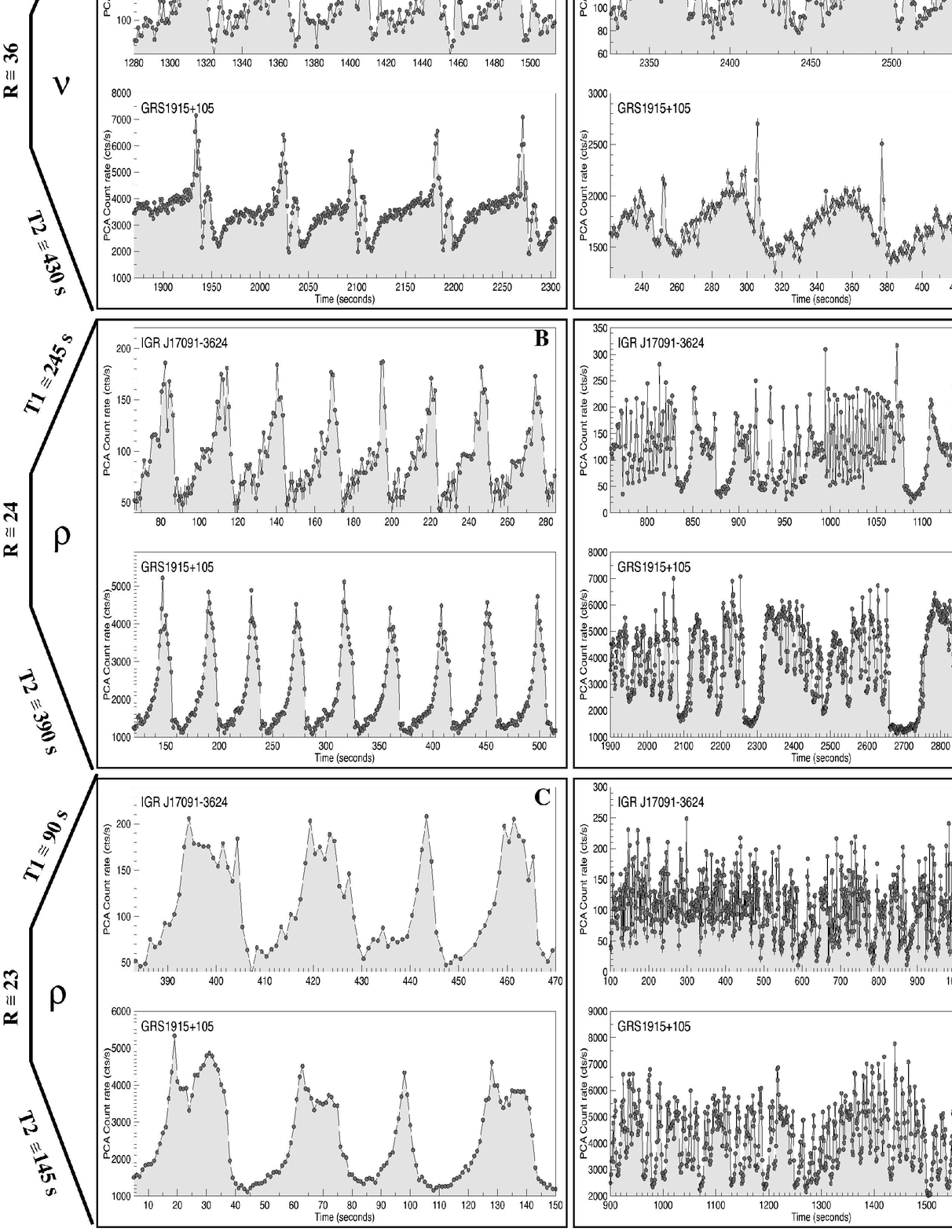}}}
\caption{Similar to Figure~\ref{fig:complc1}. The IGR~J17091--3624 D-F
  intervals come from observations 96420-04-03, -08-03 and -09-06,
  respectively.  For GRS~1915+105 they come from observations
  20187-02-01-00, 95701-01-31-00 and 10258-01-10-00, respectively. }
\label{fig:complc2}
\end{figure}

\begin{figure*} 
\centering
\resizebox{2\columnwidth}{!}{\rotatebox{0}{\includegraphics[clip]{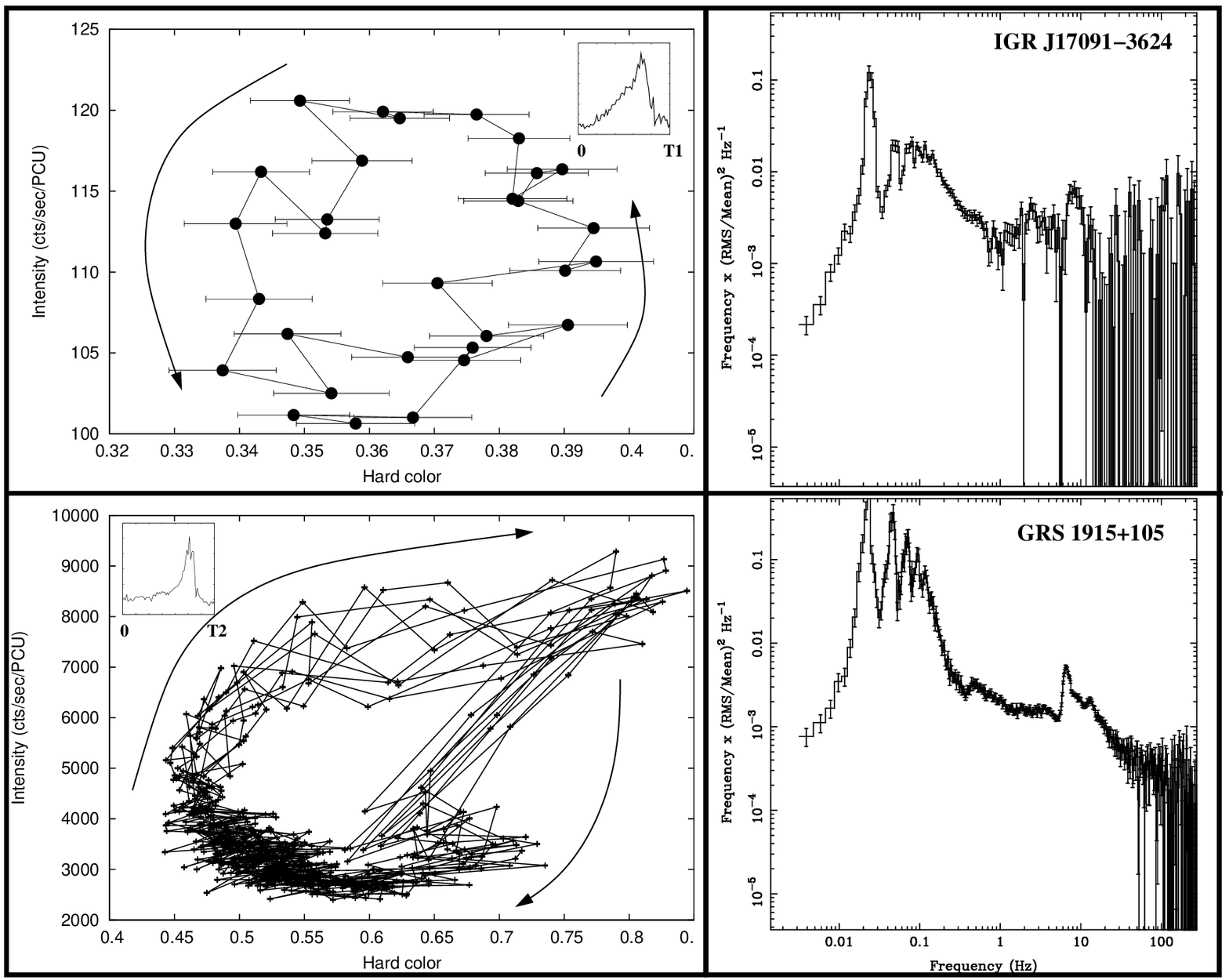}}}
\caption{\textit{Left panels} show the hardness-intensity diagram for
  flares observed during the $\rho$ variability class in
  IGR~J17091--3624 (top; ObsID 96420-01-04-03) and GRS 1915+105
  (bottom; ObsID 96378-01-01-00) occurring at an average period of
  T1$=$70.96 seconds and T2$=$63.72 seconds, respectively. Arrows mark
  the time evolution. Inset shows representative flares.  Light curves
  and colors are estimated from 1 sec averages. Intensity is the count
  rate in the 2-60 keV range (absolute channels 0-240) and hard color
  is the 6.5--15.0~keV / 2--6.5~keV count rate ratio (channels 15-35
  and 0-14, respectively).  \textit{Right panels} show representative
  power spectra from averages of 512 sec segments during the $\rho$
  variability class for IGR~J17091--364 (top, ObsID:96420-01-05-000,
  MJD 55647.9) and for GRS~1915+105 (bottom, ObsID:40703-01-07-00, MJD
  51235.3).}
\label{fig:loopPDS}
\end{figure*}

\section{Observations, data analysis}\label{sec:observations}

IGR~J17091--3624 is observed with the Proportional Counter Array
\citep[PCA; ][]{Zhang93,Jahoda06} on-board RXTE almost daily since the
outburst began in February 2011 (at the time of submission of this
Letter, IGR~J17091--3624 was still active). We use the first 147
observations, covering $\sim$180 days.
We also used 1700 archival RXTE observations of GRS~1915+105.

Power spectra and light curves were produced from the PCA using
standard techniques \citep[e.g.,][]{Belloni00, Altamirano08}.
Deadtime corrected energy spectra averaged over each observation were
created from PCU 2 data; a systematic error of 1\% was added to all
channels.
Response matrices were created with {\tt PCARSP (V11.7.1)} using the
position reported by \citet{Kennea07}, thereby taking into account a
constant 0.4$^\circ$ pointing offset used after February 23rd, 2011
(MJD 55615) to avoid a bright nearby source (see below).
All spectra were satisfactorily fitted in the 3-25 keV band with an
absorbed disk-blackbody plus power law model ({\tt phabs*(diskbb +
  powerlaw})) using Xspec 12 \citep{Arnaud96} with $N_h$ fixed to $1.1
\times 10^{22}$ cm$^{-2}$ \citep{Krimm11}; some spectra required a
Gaussian at 6.5 keV for a good fit.

\section{Results}\label{sec:results}

Figure~\ref{fig:outburstlc} (upper panel) shows the Swift/BAT
\citep{Barthelmy05} daily light curve of the first 200 days of the
outburst (assuming outburst onset on MJD $\sim$55580). The outburst
reached a maximum of 0.026 counts/cm$^2$/s (15-50 keV) $\sim$27 days
after onset; then decreased and remained below 0.01 counts/cm$^2$/s.
The lower panel of Figure~\ref{fig:outburstlc} shows the 2-20 keV,
Crab-normalized intensity and the 2-50 keV un-absorbed flux from each
RXTE/PCA pointed observations. RXTE observations started 21 days after
onset. The first 10 observations (until MJD 55615) are affected by the
contribution of the bright, variable source GX~349+2 (Sco~X-2) which
was within the 1$^\circ$ PCA FoV \citep[see,
also,][]{Rodriguez11a}. For reference, in Figure~\ref{fig:outburstlc}
we show these observations at an arbitrarily fixed intensity of 65
mCrab; we do not report their X-ray flux.

From the variability point of view, we find that during
MJD~55601--55632 , the X-ray light curves were well characterized by
broad band noise and a QPO moving in time from 0.1 Hz up to
$\sim$6~Hz.
The evolution of the power spectral components \citep[see,][for more
details]{Pahari11, Shaposhnikov11, Rodriguez11a} is similar to that
seen in BHC transients in the hard state \citep[e.g., ][]{Belloni10b},
further suggesting the black hole nature of IGR~J17091--3624.

On MJD 55634 IGR~J17091--3624 first showed a $\sim$10 mHz QPO. Since
then, a broad variety of complex patterns occurred, including strong
variability in the form of flares alternating with quiet intervals
\citep{Altamirano11b,Altamirano11c}.
The variability shows remarkable similarities with that of
GRS~1915+105 \citep[e.g.,][]{Belloni00} suggesting a common
mechanism.
In Figures~\ref{fig:complc1} and~\ref{fig:complc2} we show six
different examples; top and bottom frames of each panel correspond to
data from IGR~J17091--3624 and GRS~1915+105, respectively.

Based on the classification of \citet{Belloni00}, panel A shows
segments of the $\nu$ class, and panels B and C two different
varieties of $\rho$ class.
The $\nu$ and $\rho$ variability classes in GRS~1915+105 are
characterized by quasi-periodic ``flares'' recurring on time scales
between $\sim$40 \citep[e.g.,][]{Massaro10, Neilsen11} and $\sim$120
seconds \citep{Belloni00}.
The main differences between the $\nu$ and $\rho$ classes are that (i)
the shape and period of the flares in $\nu$ can be more irregular than
in $\rho$ and (ii) that the $\nu$ flares show a characteristic
structure in their profiles, notably a dip followed by a secondary
peak after the main one (see Figure~\ref{fig:complc1}).

Panel D shows segments of the $\alpha$ class, in which
``rounded-bumps'' are sometimes accompanied by sharp peaks which last
a few seconds. Note that in our observations of IGR~J17091--3624 we do
not see the typical $\sim$1000 s quiet interval which precedes the
bumps in the $\alpha$ class of GRS~1915+105 \citep[not shown, see,
e.g., Figure~19d in][]{Belloni00}.
Panel E shows light curves from either the $\beta$ or the $\lambda$
class. These light curves consist of quasi-periodic alternation of
low-quiet periods with highly-variable and oscillating ones.
Panel F shows segments of the $\mu$ class, where we observe periods of
very rapid (factor of 2-4 changes in count rate in less than 5
seconds) and almost incoherent variability.

In addition to the light curves shown in Figures~\ref{fig:complc1}
and~\ref{fig:complc2} , IGR~J17091--3624 displays complex light curves
combining characteristics of different classes.
It also shows \textit{quiet periods} in which 1-sec light curves are
flat ($<10$\% fractional rms amplitude at $<1$~Hz). These periods are
marked with grey shadowed areas in Figure~\ref{fig:outburstlc} and
coincide with the times when IGR~J17091-3624 is detected most strongly
by \textit{Swift/BAT}, i.e. when the spectrum is hard. In terms of the
GRS~1915+105 variability classes, these periods could correspond to
the $\chi$ class.

Figures~\ref{fig:complc1} and~\ref{fig:complc2} not only shows the
remarkable similarities in the shapes of light curves from both
sources, it also shows two clear differences:
(i) the time scales can be different (IGR~J17091--3624 tends to be
faster, see below) and
(ii) the average count rate (or flux) of the source can be much higher
(factor 10-50) in GRS~1915+105.
Some of the spectral and timing analysis performed for GRS~1915+105 on
time scales of seconds \citep[e.g.,][]{Markwardt99, Belloni00,
  Done04,Soleri08} is prevented by the combination of relatively low
count rate and faster variability in IGR~J17091--3624.

Given that we find the $\nu$ and $\rho$ variability classes (panels
A-C) in about 35\% of our observations (another $\sim$35\% of the
light curves are flat, and the remaining $\sim$30\% are a mix of the
$\alpha$, $\beta$, $\mu$, $\lambda$ and unclassified ones) and that
$\rho$ is one of the best studied classes in GRS~1915+105, in the rest
of this Letter we constrain ourselves on further comparison of the
$\nu$ and $\rho$ classes between sources.
More detailed comparison of the other variability classes will be
presented in upcoming papers.

Some of the observations clearly show only the $\rho$ or the $\nu$
types of flares, some show a mix and sometimes differentiating the two
classes is difficult due to the low statistics.
In any case, these flares can occur as fast as every few (2-5) seconds
(e.g., ObsID:~96420-01-22-04, MJD~55768), and as slow as every
$\sim$100 seconds (e.g., ObsID~96420-01-03-01, MJD~55634).
This means that the oscillations in IGR~J17091--3624 can be faster
than those in GRS~1915+105, but not as slow.
The fractional rms amplitude of the flares covers a range from
$\sim$2\% to up to 50-60\%.
If one assumes that the minimum period that a quasi-periodic feature
can reach scales proportional to some power of the mass of the compact
object \citep[see, e.g.,][]{Belloni97,Frank02}, and that a 2-5 s
recurrence time of the $\rho$/$\nu$ flares IGR~J17091--3624 versus
$\sim$40 s in GRS~1915+105 is due to a difference in mass, then our
results suggest the black hole in IGR~J17091--3624 could be a factor
of a few less massive than the $14\pm4.4 \ M_{\odot}$ of GRS~1915+105.

The variability classes are known to exhibit distinctive spectral
evolution \citep[e.g.,][]{Belloni00}. In a color-color diagram (CD) or
hardness-intensity diagram (HID), one observes that each flare from
the $\rho$ (and sometimes $\nu$) class traces a loop \citep[or
``ring'', see][]{Vilhu98,Belloni00}. Figure~\ref{fig:loopPDS}
(bottom-left) shows the HID for ten consecutive flares from a single
observation of GRS~1915+105 (inset shows a representative flare).
The loop in the HID is always traversed clockwise.

The HIDs and CDs from \textit{single} flares of IGR~J17091--3624 are
dominated by low statistics, so we could neither exclude nor confirm
loop-like patterns similar to those observed in GRS~1915+105.  
To improve statistics, we used intervals where more than 10 flares
occurred approximately periodically and folded each interval at the
best average period.
Although this process washed out some of the structure in the light
curves (due to the quasi-periodicity of the signal and the profile
differences between flares), all the HIDs we created using the same
energy bands as in GRS~1915+105 (which are fixed by the observing
modes used) always showed a loop resembling that seen in
GRS~1915+105. In the upper-left panel of Figure~\ref{fig:loopPDS} we
show a representative example (inset shows the average profile of the
flare). However, in all cases the loop is traversed in an
anti-clockwise sense, i.e., opposite to what we see in GRS~1915+105.
As the hardness ratios are a crude characterization of the spectrum,
detailed spectral modeling of these loops \citep[e.g.,
][]{Neilsen11} are needed to understand whether there is a physical
difference between the $\rho$ class seen in both sources.

Figure~\ref{fig:loopPDS} (right panels) shows representative power
spectra of the $\rho$ variability class in IGR~J17091--3624 and
GRS~1915+105 .
The power spectra share the same main features:
(i) a low-frequency QPO (with high harmonic content) due to the
``flares'' and
(ii) a QPO with characteristic frequency between 6-10 Hz \citep[e.g.,
][]{Muno99}.
In addition, for both sources we sometimes find a ``bump'' with
characteristic frequency between 1 and 5 Hz.

\section{Discussion}

An extensive literature exists attempting to understand the complex
variability observed in GRS~1915+105.  Some authors propose that the
high luminosity (close to, or super-Eddington ) of GRS~1915+105 is the
determining factor \citep[e.g.,][and references therein]{Belloni97,
  Vilhu98, Belloni00, Nayakshin00, Janiuk02, Done04, Neilsen11}.
Although it is generally accepted that the complex X-ray variability
in GRS~1915+105 results from disk instabilities, the exact nature of
the instability remains unknown.  The lack of at least a second source
showing similar characteristics has prevented definite conclusions.

In this Letter we show for the first time that another source, the BHC
IGR~J17091-3624, can show the same broad variety of complex light
curves as GRS~1915+105 (at least 7 of the 12 variability classes
observed in GRS~1915+105, in addition to unclassified ones; note that
at the time of submission of this Letter IGR~J17091-3624 is still
active).
Although the comparisons presented in this paper constitute only a
first step, the observed similarities suggest that the complex light
curves of the two sources are produced by the same physical
mechanisms.
If true, the low flux of IGR~J17091-3624 compared with GRS~1915+105
combined with the circumstance that currently neither the distance to
IGR~J17091-3624 nor the mass of its compact object are known, raises
the fundamental question: is IGR~J17091-3624 close to Eddington or not
at times when showing the same characteristic X-ray variability?

A scenario in which IGR~J17091-3624 is not emitting at close to the
Eddington but only at a few percent, is at variance with models where
the variability is explained as due to disk instabilities that
\textit{only} occur at high luminosity \citep[e.g.,][and references
therein]{Vilhu98, Belloni00, Nayakshin00, Janiuk02, Done04,
  Neilsen11}. However, in this scenario IGR~J17091-3624 would follow
radio/X-ray correlation of some or most BHCs \citep[depending if the
distance is closer to $\sim11$ kpc or $\sim17$ kpc,
respectively;][]{Rodriguez11}.

\begin{figure} 
\centering
\resizebox{1\columnwidth}{!}{\rotatebox{-90}{\includegraphics[clip]{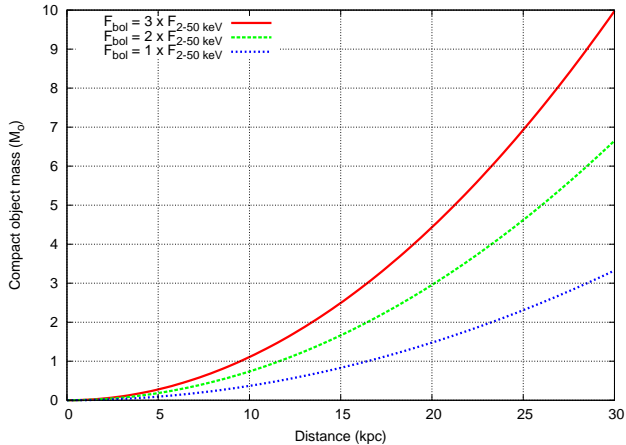}}}
\caption{Mass of the compact object versus distance to the binary
  system assuming IGR~J17091-3624 is emitting at Eddington. The three
  curves correspond to different (1, 2 and 3) correction factors
  between the 2-50 keV and the bolometric flux. }
\label{fig:edd}
\end{figure}

A scenario in which IGR~J17091-3624 is emitting at close to the
Eddington limit puts constraints on its mass and distance.
Although this scenario would imply that, independently of the
distance, IGR~J17091-3624 does not follow the standard Radio/X-ray
luminosity plane for BHCs \citep{Rodriguez11}, this would be similar
to GRS~1915+105, as it does not follow the radio/X-ray flux plane
either \citep[see, e.g., figure 4 in ][ and references
therein]{Rodriguez11}.

We therefore made a rough estimate of the source flux in each
observation by calculating the average-per-observation energy
spectrum. In the upper panel of Figure~\ref{fig:outburstlc} we plot
the 2-50 keV unabsorbed flux.
In IGR~J17091-3624 we observe the $\alpha$, $\beta$, $\nu$, $\rho$ and
$\mu$ classes when the flux is between $\sim$2 and $\sim$3.7 $10^{-9}$
erg~cm$^{-2}$~s$^{-1}$ (the {\tt diskbb} temperature {\tt kT} vary
between 1 and 2 keV while the power law index between 2 and 3). 
Assuming that IGR~J17091-3624 is emitting at Eddington rates, we
derived the mass of the compact object as a function of distance
(Figure~\ref{fig:edd}) for a 2-50 keV flux of $4 \cdot
10^{-9}$~erg~cm$^{-2}$~s$^{-1}$.
Given that the correction factor linking the 2-50 keV and the
bolometric flux is not exactly known, we plot curves for factors of 1,
2 and 3.
Figure~\ref{fig:edd} implies that if IGR~J17091-3624 emits at
Eddington, then either it harbors the lowest mass black hole known
today ($<3 M_\odot$ for distances lower than 17 kpc), or, it is very
distant. Such a large distance, together with its
$b\simeq$2.2$^\circ$ Galactic latitude, would imply a significant, but
not necessarily implausible, altitude above the disk \citep[e.g.,
$b\simeq$-2.8$^\circ$ and $d>25$kpc for the BHC
GS~1354--64,][]{Casares09}.

Clearly, constraining the distance to IGR~J17091-3624 is fundamental
for understanding the physical processes that govern the X-ray
variability in IGR~J17091-3624, and hence those in GRS~1915+105.
The interstellar absorption in the direction of IGR~J17091-3624 is
high \citep[$\sim1 \times 10^{22}$ cm$^{-2}$, see][]{Rodriguez11},
implying that optical studies when the binary system is in quiescence
\citep[e.g., ][]{Casares04} might be challenging.
However, accurate parallax distance estimates like those reported for
the black hole X-ray binary V404 Cyg \citep{Miller-Jones09} could
still be possible.

\textbf{Acknowledgments:} We thank Dave Russell and JamesMiller-Jones
for insightful discussions. M.L. acknowledges support from an NWO
Rubicon fellowship. TB has received funding from the European
Community’s Seventh Framework Programme (FP7/2007-2013) under grant
agreement number ITN 215212 “Black Hole Universe,” and TMD from the
Spanish MEC under theConsolider-Ingenio 2010 Programme grant CSD2006-
00070: “First Science with the GTC.”


\begin{thebibliography}{48}
\expandafter\ifx\csname natexlab\endcsname\relax\def\natexlab#1{#1}\fi

\bibitem[{{Altamirano} {et~al.}(2008){Altamirano}, {van der Klis},
  {M{\'e}ndez}, {Jonker}, {Klein-Wolt}, \& {Lewin}}]{Altamirano08}
{Altamirano}, D., {van der Klis}, M., {M{\'e}ndez}, M., {Jonker}, P.~G.,
  {Klein-Wolt}, M., \& {Lewin}, W.~H.~G. 2008, \apj, 685, 436

\bibitem[{{Altamirano} {et~al.}(2011{\natexlab{a}}){Altamirano}, {Linares},
  {van der Klis}, {Wijnands}, {Kalamkar}, {Casella}, {Watts}, {Patruno},
  {Armas-Padilla}, {Cavecchi}, {Degenaar}, {Kaur}, {Yang}, \&
  {Rea}}]{Altamirano11a}
{Altamirano}, D., {et~al.} 2011{\natexlab{a}}, The Astronomer's Telegram, 3225

\bibitem[{{Altamirano} {et~al.}(2011{\natexlab{b}}){Altamirano}, {Belloni},
  {Krimm}, {Casella}, {Curran}, {Kennea}, {Kalamkar}, {van der Klis},
  {Wijnands}, {Linares}, {Motta}, {Munoz-Darias}, \& {Stiele}}]{Altamirano11b}
---. 2011{\natexlab{b}}, The Astronomer's Telegram, 3230

\bibitem[{{Altamirano} {et~al.}(2011{\natexlab{c}}){Altamirano}, {Belloni},
  {Krimm}, {Casella}, {Curran}, {Kennea}, {Kalamkar}, {van der Klis},
  {Wijnands}, {Linares}, {Motta}, {Munoz-Darias}, \& {Stiele}}]{Altamirano11c}
---. 2011{\natexlab{c}}, The Astronomer's Telegram, 3299

\bibitem[{{Arnaud}(1996)}]{Arnaud96}
{Arnaud}, K.~A. 1996, in Astronomical Society of the Pacific Conference Series,
  Vol. 101, Astronomical Data Analysis Software and Systems V, ed. G.~H.
  {Jacoby} \& J.~{Barnes}, 17

\bibitem[{{Barthelmy} {et~al.}(2005){Barthelmy}, {Barbier}, {Cummings},
  {Fenimore}, {Gehrels}, {Hullinger}, {Krimm}, {Markwardt}, {Palmer},
  {Parsons}, {Sato}, {Suzuki}, {Takahashi}, {Tashiro}, \&
  {Tueller}}]{Barthelmy05}
{Barthelmy}, S.~D., {et~al.} 2005, Space Science Reviews, 120, 143

\bibitem[{{Belloni} {et~al.}(2000){Belloni}, {Klein-Wolt}, {M{\'e}ndez}, {van
  der Klis}, \& {van Paradijs}}]{Belloni00}
{Belloni}, T., {Klein-Wolt}, M., {M{\'e}ndez}, M., {van der Klis}, M., \& {van
  Paradijs}, J. 2000, \aap, 355, 271

\bibitem[{{Belloni} {et~al.}(1997){Belloni}, {Mendez}, {King}, {van der Klis},
  \& {van Paradijs}}]{Belloni97}
{Belloni}, T., {Mendez}, M., {King}, A.~R., {van der Klis}, M., \& {van
  Paradijs}, J. 1997, \apjl, 479, L145

\bibitem[{{Belloni}(2010)}]{Belloni10b}
{Belloni}, T.~M. 2010, in Lecture Notes in Physics, Berlin Springer Verlag,
  Vol. 794, Lecture Notes in Physics, Berlin Springer Verlag, ed. {T.~Belloni},
  53

\bibitem[{{Belloni} {et~al.}(2011){Belloni}, {Motta}, \&
  {Mu{\~n}oz-Darias}}]{Belloni11}
{Belloni}, T.~M., {Motta}, S.~E., \& {Mu{\~n}oz-Darias}, T. 2011, eprint
  arXiv:1109.3388

\bibitem[{{Capitanio} {et~al.}(2006){Capitanio}, {Bazzano}, {Ubertini},
  {Zdziarski}, {Bird}, {De Cesare}, {Dean}, {Stephen}, \&
  {Tarana}}]{Capitanio06}
{Capitanio}, F., {et~al.} 2006, \apj, 643, 376

\bibitem[{{Capitanio} {et~al.}(2009){Capitanio}, {Giroletti}, {Molina},
  {Bazzano}, {Tarana}, {Kennea}, {Dean}, {Hill}, {Tavani}, \&
  {Ubertini}}]{Capitanio09}
---. 2009, \apj, 690, 1621

\bibitem[{{Casares} {et~al.}(2004){Casares}, {Zurita}, {Shahbaz}, {Charles}, \&
  {Fender}}]{Casares04}
{Casares}, J., {Zurita}, C., {Shahbaz}, T., {Charles}, P.~A., \& {Fender},
  R.~P. 2004, \apjl, 613, L133

\bibitem[{{Casares} {et~al.}(2009){Casares}, {Orosz}, {Zurita}, {Shahbaz},
  {Corral-Santana}, {McClintock}, {Garcia}, {Mart{\'{\i}}nez-Pais}, {Charles},
  {Fender}, \& {Remillard}}]{Casares09}
{Casares}, J., {et~al.} 2009, \apjs, 181, 238

\bibitem[{{Castro-Tirado} {et~al.}(1992){Castro-Tirado}, {Brandt}, \&
  {Lund}}]{Castro-Tirado92}
{Castro-Tirado}, A.~J., {Brandt}, S., \& {Lund}, N. 1992, \iaucirc, 5590, 2

\bibitem[{{Done} {et~al.}(2004){Done}, {Wardzi{\'n}ski}, \&
  {Gierli{\'n}ski}}]{Done04}
{Done}, C., {Wardzi{\'n}ski}, G., \& {Gierli{\'n}ski}, M. 2004, \mnras, 349,
  393

\bibitem[{{Fender} \& {Belloni}(2004)}]{Fender04}
{Fender}, R., \& {Belloni}, T. 2004, \araa, 42, 317

\bibitem[{{Frank} {et~al.}(2002){Frank}, {King}, \& {Raine}}]{Frank02}
{Frank}, J., {King}, A., \& {Raine}, D.~J. 2002, {Accretion Power in
  Astrophysics: Third Edition}, ed. {UK: Cambridge University Press}

\bibitem[{{Greiner} {et~al.}(2001){Greiner}, {Cuby}, \&
  {McCaughrean}}]{Greiner01}
{Greiner}, J., {Cuby}, J.~G., \& {McCaughrean}, M.~J. 2001, \nat, 414, 522

\bibitem[{{Harlaftis} \& {Greiner}(2004)}]{Harlaftis04}
{Harlaftis}, E.~T., \& {Greiner}, J. 2004, \aap, 414, L13

\bibitem[{{in't Zand} {et~al.}(2003){in't Zand}, {Heise}, {Lowes}, \&
  {Ubertini}}]{Intzand03}
{in't Zand}, J.~J.~M., {Heise}, J., {Lowes}, P., \& {Ubertini}, P. 2003, The
  Astronomer's Telegram, 160

\bibitem[{{Jahoda} {et~al.}(2006){Jahoda}, {Markwardt}, {Radeva}, {Rots},
  {Stark}, {Swank}, {Strohmayer}, \& {Zhang}}]{Jahoda06}
{Jahoda}, K., {Markwardt}, C.~B., {Radeva}, Y., {Rots}, A.~H., {Stark}, M.~J.,
  {Swank}, J.~H., {Strohmayer}, T.~E., \& {Zhang}, W. 2006, \apjs, 163, 401

\bibitem[{{Janiuk} {et~al.}(2002){Janiuk}, {Czerny}, \&
  {Siemiginowska}}]{Janiuk02}
{Janiuk}, A., {Czerny}, B., \& {Siemiginowska}, A. 2002, \apj, 576, 908

\bibitem[{{Kennea} \& {Capitanio}(2007)}]{Kennea07}
{Kennea}, J.~A., \& {Capitanio}, F. 2007, The Astronomer's Telegram, 1140

\bibitem[{{Krimm} {et~al.}(2011){Krimm}, {Barthelmy}, {Baumgartner},
  {Cummings}, {Fenimore}, {Gehrels}, {Kennea}, {Markwardt}, {Palmer},
  {Sakamoto}, {Skinner}, {Stamatikos}, {Tueller}, \& {Ukwatta}}]{Krimm11}
{Krimm}, H.~A., {et~al.} 2011, The Astronomer's Telegram, 3144

\bibitem[{{Kuulkers} {et~al.}(2003){Kuulkers}, {Lutovinov}, {Parmar},
  {Capitanio}, {Mowlavi}, \& {Hermsen}}]{Kuulkers03b}
{Kuulkers}, E., {Lutovinov}, A., {Parmar}, A., {Capitanio}, F., {Mowlavi}, N.,
  \& {Hermsen}, W. 2003, The Astronomer's Telegram, 149

\bibitem[{{Markwardt} {et~al.}(1999){Markwardt}, {Swank}, \&
  {Taam}}]{Markwardt99}
{Markwardt}, C.~B., {Swank}, J.~H., \& {Taam}, R.~E. 1999, \apjl, 513, L37

\bibitem[{{Massaro} {et~al.}(2010){Massaro}, {Ventura}, {Massa}, {Feroci},
  {Mineo}, {Cusumano}, {Casella}, \& {Belloni}}]{Massaro10}
{Massaro}, E., {Ventura}, G., {Massa}, F., {Feroci}, M., {Mineo}, T.,
  {Cusumano}, G., {Casella}, P., \& {Belloni}, T. 2010, \aap, 513, A21

\bibitem[{{Miller-Jones} {et~al.}(2009){Miller-Jones}, {Jonker}, {Dhawan},
  {Brisken}, {Rupen}, {Nelemans}, \& {Gallo}}]{Miller-Jones09}
{Miller-Jones}, J.~C.~A., {Jonker}, P.~G., {Dhawan}, V., {Brisken}, W.,
  {Rupen}, M.~P., {Nelemans}, G., \& {Gallo}, E. 2009, \apjl, 706, L230

\bibitem[{{Mirabel} {et~al.}(1998){Mirabel}, {Dhawan}, {Chaty}, {Rodriguez},
  {Marti}, {Robinson}, {Swank}, \& {Geballe}}]{Mirabel98}
{Mirabel}, I.~F., {Dhawan}, V., {Chaty}, S., {Rodriguez}, L.~F., {Marti}, J.,
  {Robinson}, C.~R., {Swank}, J., \& {Geballe}, T. 1998, \aap, 330, L9

\bibitem[{{Mirabel} \& {Rodr{\'{\i}}guez}(1994)}]{Mirabel94}
{Mirabel}, I.~F., \& {Rodr{\'{\i}}guez}, L.~F. 1994, \nat, 371, 46

\bibitem[{{Morgan} {et~al.}(1997){Morgan}, {Remillard}, \&
  {Greiner}}]{Morgan97}
{Morgan}, E.~H., {Remillard}, R.~A., \& {Greiner}, J. 1997, \apj, 482, 993

\bibitem[{{Muno} {et~al.}(1999){Muno}, {Morgan}, \& {Remillard}}]{Muno99}
{Muno}, M.~P., {Morgan}, E.~H., \& {Remillard}, R.~A. 1999, \apj, 527, 321

\bibitem[{{Nayakshin} {et~al.}(2000){Nayakshin}, {Rappaport}, \&
  {Melia}}]{Nayakshin00}
{Nayakshin}, S., {Rappaport}, S., \& {Melia}, F. 2000, \apj, 535, 798

\bibitem[{{Neilsen} {et~al.}(2011){Neilsen}, {Remillard}, \& {Lee}}]{Neilsen11}
{Neilsen}, J., {Remillard}, R.~A., \& {Lee}, J.~C. 2011, \apj, 737, 69

\bibitem[{{Pahari} {et~al.}(2011){Pahari}, {Yadav}, \&
  {Bhattacharyya}}]{Pahari11}
{Pahari}, M., {Yadav}, J., \& {Bhattacharyya}, S. 2011, ApJ Submitted,
  astro-ph:1105.4694

\bibitem[{{Reig} {et~al.}(2000){Reig}, {Belloni}, {van der Klis}, {M{\'e}ndez},
  {Kylafis}, \& {Ford}}]{Reig00a}
{Reig}, P., {Belloni}, T., {van der Klis}, M., {M{\'e}ndez}, M., {Kylafis},
  N.~D., \& {Ford}, E.~C. 2000, \apj, 541, 883

\bibitem[{{Remillard} \& {McClintock}(2006)}]{Remillard06}
{Remillard}, R.~A., \& {McClintock}, J.~E. 2006, \araa, 44, 49

\bibitem[{{Revnivtsev} {et~al.}(2003){Revnivtsev}, {Gilfanov}, {Churazov}, \&
  {Sunyaev}}]{Revnivtsev03a}
{Revnivtsev}, M., {Gilfanov}, M., {Churazov}, E., \& {Sunyaev}, R. 2003, The
  Astronomer's Telegram, 150

\bibitem[{{Rodriguez} {et~al.}(2011{\natexlab{a}}){Rodriguez}, {Corbel},
  {Caballero}, {Tomsick}, {Tzioumis}, {Paizis}, {Cadolle Bel}, \&
  {Kuulkers}}]{Rodriguez11a}
{Rodriguez}, J., {Corbel}, S., {Caballero}, I., {Tomsick}, J.~A., {Tzioumis},
  T., {Paizis}, A., {Cadolle Bel}, M., \& {Kuulkers}, E. 2011{\natexlab{a}},
  \aap, 533, L4

\bibitem[{{Rodriguez} {et~al.}(2011{\natexlab{b}}){Rodriguez}, {Corbel},
  {Tomsick}, {Paizis}, \& {Kuulkers}}]{Rodriguez11}
{Rodriguez}, J., {Corbel}, S., {Tomsick}, J.~A., {Paizis}, A., \& {Kuulkers},
  E. 2011{\natexlab{b}}, The Astronomer's Telegram, 3168

\bibitem[{{Shaposhnikov}(2011)}]{Shaposhnikov11}
{Shaposhnikov}, N. 2011, The Astronomer's Telegram, 3179

\bibitem[{{Soleri} {et~al.}(2008){Soleri}, {Belloni}, \& {Casella}}]{Soleri08}
{Soleri}, P., {Belloni}, T., \& {Casella}, P. 2008, \mnras, 383, 1089

\bibitem[{{Strohmayer}(2001)}]{Strohmayer01b}
{Strohmayer}, T.~E. 2001, \apjl, 554, L169

\bibitem[{{Tagger} {et~al.}(2004){Tagger}, {Varni{\`e}re}, {Rodriguez}, \&
  {Pellat}}]{Tagger04}
{Tagger}, M., {Varni{\`e}re}, P., {Rodriguez}, J., \& {Pellat}, R. 2004, \apj,
  607, 410

\bibitem[{{van der Klis}(2006)}]{Vanderklis06}
{van der Klis}, M. 2006, in Compact Stellar X-Ray Sources, ed. W. H. G. Lewin
  \& M. van der Klis (Cambridge: Cambridge Univ. Press)

\bibitem[{{Vilhu} \& {Nevalainen}(1998)}]{Vilhu98}
{Vilhu}, O., \& {Nevalainen}, J. 1998, \apjl, 508, L85

\bibitem[{{Zhang} {et~al.}(1993){Zhang}, {Giles}, {Jahoda}, {Soong}, {Swank},
  \& {Morgan}}]{Zhang93}
{Zhang}, W., {Giles}, A.~B., {Jahoda}, K., {Soong}, Y., {Swank}, J.~H., \&
  {Morgan}, E.~H. 1993, in Proc. SPIE Vol. 2006, p. 324-333, EUV, X-Ray, and
  Gamma-Ray Instrumentation for Astronomy IV, Oswald H. Siegmund; Ed., 324--333

\end{thebibliography}
\end{document}